\documentclass[12pt,preprint]{aastex}
\usepackage{graphicx}
\usepackage{epsfig}
\usepackage{natbib}
\usepackage{amsmath, amsthm, amssymb}
\usepackage{longtable}
\usepackage{lipsum}

\title{Universal Behavior of X-ray Flares from Black Hole Systems}
\author
{F. Y. Wang$^{1,2}$, Z. G. Dai$^{1,2}$, S. X. Yi$^{1,2}$ \& S. Q.
Xi$^3$
\\
$^1$School of Astronomy and Space Science, Nanjing University,
Nanjing 210093, China\\
$^2$Key Laboratory of Modern Astronomy and Astrophysics (Nanjing
University), Ministry of Education, Nanjing 210093, China\\
$^3$ Department of Physics and GXU-NAOC Center for Astrophysics and
Space Sciences, Guangxi University, Nanning 530004, China}

\begin{document}
\altaffiltext{}{E-mail: fayinwang@nju.edu.cn; dzg@nju.edu.cn}

\begin{abstract}
X-ray flares have been discovered in black hole systems, such as
gamma-ray bursts, the tidal disruption event Swift J1644+57, the
supermassive black hole Sagittarius A$^*$ at the center of our
Galaxy, and some active galactic nuclei. Their occurrences are
always companied by relativistic jets. However, it is still unknown
whether there is a physical analogy among such X-ray flares produced
in black hole systems spanning nine orders of magnitude in mass.
Here we report the observed data of X-ray flares, and show that they
have three statistical properties similar to solar flares, including
power-law distributions of energies, durations, and waiting times,
which both can be explained by a fractal-diffusive self-organized
criticality model. These statistical similarities, together with the
fact that solar flares are triggered by a magnetic reconnection
process, suggest that all of the X-ray flares are consistent with
magnetic reconnection events, implying that their concomitant
relativistic jets may be magnetically dominated.
\end{abstract}


\vspace{1.0cm}

\section{Introduction}
X-ray flares are common astrophysical explosive phenomena throughout
the universe. They have been observed in stars, stellar-mass black
holes, and supermassive black holes (SMBHs) located at the center of
galaxies, particularly in the Sun \citep{shi11}, gamma-ray bursts
(GRBs) \citep{geh09}, the tidal disruption event Swift J1644+57
\citep{bur11,blo11}, the SMBH Sagittarius A$^*$ (named Sgr A$^*$) at
the center of our Galaxy \citep{bag01}, and some active galactic
nuclei (AGN) \citep{ree84}. These black hole systems, spanning nine
orders of magnitude in mass, always generate relativistic jets (De
Young 1991; Mirabel \& Rodriguez 1999; Meier et al. 2001; Zhang
2007), which are likely to produce X-ray flares with short rise and
decay times. Until now, the composition of such relativistic jets
has been poorly known (Harris \& Krawczynski 2006). Meanwhile the
physical origin of resultant X-ray flares has also remained
mysterious, although some models have been proposed for X-ray flares
in GRBs \citep{dai06,mes06}, Swift J1644+57 \citep{wan12}, Sgr A$^*$
\citep{mar01,yua03,yua04,dod10}, and M87 \citep{har03}. It is well
known that solar flares originate from magnetic reconnection
\citep{shi11,zwe09}. Here we investigate the energy frequency
distribution, duration-time distribution and waiting time
distribution of X-ray flares from black hole systems, compare these
distributions with those of solar flares, and infer the physical
origin of X-ray flares and the composition of relativistic jets.

In astrophysics, there is consensus about a common set of
morphological properties, which may be shared by most of X-ray
flares spanning many orders of magnitude in power. Interestingly,
there is evidence that GRB X-ray flares and solar flares have
similar statistical properties, suggesting a similar physical
origin, i.e. magnetic reconnection \citep{wan13}. However, a clear
connection among X-ray flares from relativistic jets in black hole
systems with vastly different masses has not yet been built, though
a recent study shows that relativistic jets in GRBs and AGNs may
have a similar efficiency of energy dissipation \citep{nem12}. Wang
et al. (2014) found that the correlation between radio luminosity
and peak energy of GRBs is similar as that of blazars, which
indicated that the radiation process of GRBs is synchrotron
radiation both in the prompt and afterglow phases.

Solar flares are believed to be self-organized criticality (SOC)
events driven by magnetic reconnection
\citep{lu91,cha01,mor08,asc12}. The concept of SOC was first
proposed by Bak et al. (1987, 1988), which has been initially
applied to sandpile avalanches, and has been generalized to
nonlinear dissipative systems that are driven in a critical state.
Wang \& Dai (2013) found that X-ray flares from GRBs show SOC
behaviors. Whether the SOC characteristics exist in X-ray flares in
black hole systems except for GRBs remains unclear, though some hints have been
found \citep{min94,tak95,neg95,cip03,zha07}.

In this paper, we study the energy frequency distribution,
duration-time distribution and waiting time distribution of X-ray
flares from black hole systems, such as Swift J1644+57, Sgr A$^*$
and M87. The structure of this paper is arranged as follows. In
section 2, we derive the X-ray flare data. We present the fitting
results and explanations in section 3. The conclusions and
discussion are given in section 4.

\section{The X-ray flare data} \label{obs}
We focus on X-ray flares from SMBHs, such as the central black hole
of Swift J1644+57, Sgr A$^*$ at our Galactic center, and the AGN
M87.

\subsection{X-ray flares of Swift J1644+57}
Swift J1644+57 is the first event with a relativistic jet generated
by the tidal disruption of a star by a SMBH \citep{bur11,blo11}.
This event is at redshift $z=0.35$. The central black hole mass of
Swift J1644+57 is about $7.4\times 10^6M_\odot$ \citep{bur11}, where
$M_\odot$ is the solar mass. There are many flares presented in the
X-ray light curve of Swift J1644+57 \citep{bur11}. The X-ray light
curve of Swift J1644+57 is obtained from http://www.swift.ac.uk
/xrt\_curves \citep{eva07}, which was observed by the Swift
satellite \citep{geh04}. The X-ray flux in the 0.3-10 keV band is
shown in Figure 1. The X-ray telescope (XRT) on board Swift has a
number of different operating modes, depending on the brightness of
an observed source. The time resolution is from millisecond to
seconds \citep{geh04}. The data are binned dynamically, i.e., the
binning criteria vary with count rate \citep{eva07}. The solid line
represents the underlying continuum emission, which is dramatically
consistent with the Chandra X-ray observations in about 500 days
when the relativistic jet shuts off \citep{zau13}. When we fit the
parameters of flares, the underlying continuum emission is
subtracted. The total X-ray emission shows a constant flux at $t<15$
days \citep{bur11} and $F_X\propto t^{-5/3}$ at $t>15$ days
\citep{ber12}. The constant flux may be due to the extraction of the
rotational energy of a spinning supermassive black hole through the
Blandford-Znajek mechanism \citep{bla77,lei11}. The steady decline
$F_X\propto t^{-5/3}$ traces the mass accretion rate of
post-disruption debris \citep{ree88}. We fit the flares with a
smooth broken power-law function \citep{li12}
\begin{equation}\label{SBPL}
F(t)=F_{0}\left[\left(\frac{t}{t_{b}}\right)^{\alpha_{1}\omega}+\left(\frac{t}{t_{b}}\right)^{\alpha_{2}\omega}\right]^{-\frac{1}{\omega}},
\end{equation}
where $\alpha_{1}$ and $\alpha_{2}$ are the temporal slopes, $t_{b}$
is the break time, and $\omega$ measures the sharpness of a peak of
the light curve component. This method is very similar to the
fitting method of GRB X-ray flares \citep{chi07,chi10,fal07}. The
example of best-fit flares is shown in Figure 2. The flare
parameters including the start time, end time, fluence, fluence
error and isotropic energy are shown in Table 1. The total number of
X-ray flares of Swift J1644+57 is 68. These parameters are derived
as follows. The isotropic energy of one flare in the 0.3-10 keV band
can be calculated by $E_{\rm iso}=4\pi d^2_L(z)S_F/(1+z)$, where
$S_F$ is the fluence, $z=0.35$ is the redsihft, and $d_L(z)$ is the
luminosity distance calculated for a flat $\Lambda$CDM universe with
$\Omega_M=0.3$, $\Omega_\Lambda=0.7$ and $H_0=70$
km\,s$^{-1}$\,Mpc$^{-3}$. The error of flare energy is calculated by
$\sigma_E=4\pi d^2_L(z)\sigma_{S_F}/(1+z)$, where $\sigma_{S_F}$ is
the error of fluence. The waiting time in the source's rest frame
can be obtained by $\Delta t=(t_{i+1}-t_i)/(1+z)$, where $t_{i+1}$
is the observed start time of the $i+1$th flare, $t_i$ is the
observed start time of the $i$th flare, and $1+z$ is the factor to
transfer the time into the source-frame one. The duration time is
calculated as $T=(t_{\rm end}-t_{\rm start})/(1+z)$.

\subsection{X-ray flares of Sgr A$^*$}
The SMBH Sgr A$^*$ at our Galactic center has a mass of $4.1\times
10^6M_\odot$ at its distance of 8 kiloparsecs to the Earth
\citep{ghe08}. Sgr A$^*$ shows flares in the X-ray and near infrared
bands \citep{dod09}. The X-ray emission of Sgr A$^*$ typically has a
luminosity of a few times $10^{33}$ erg s$^{-1}$ \citep{bag03}, but
it shows rapid X-ray flaring sometimes. Here we use the
observational data from the Chandra X-ray Observatory's 2012 Sgr
A$^*$ X-ray Visionary Project \citep{wanD13}. In 3 mega-seconds of
Chandra observations, 39 X-ray flares from Sgr A$^*$ with duration
from a few hundred seconds to 8000 seconds and with 2-10 keV
luminosity from 10$^{34}$ erg s$^{-1}$ to $2\times10^{35}$ erg
s$^{-1}$ were detected \citep{nei13}. The flare parameters of Sgr
A$^*$ are compiled in Table 2, which are taken from \citet{nei13}.
The underlying emission is subtracted when the parameters of flares
are derived \citep{nei13}. The waiting time is calculated from
$\Delta t=(t_{i+1}-t_i)$, where $t_{i+1}$ is the observed start time
of the $i+1$th flare, $t_i$ is the observed start time of the $i$th
flare. Neilsen et al. (2013) found that the X-ray flare luminosity
distribution $dN/dL$ is consistent with a power law with index about
$-1.9$, and the duration time distribution is a power law with index
$-0.9\pm 0.2$.

\subsection{X-ray flares of M87}
The first jet was discovered in 1918 within the elliptical galaxy
M87 in the Virgo cluster. M87 shows X-ray flaring from nucleus,
HST-1, knot A, and knot D. The very high-energy (E$>$ 100 GeV)
flares have been detected. In this study, we use the X-ray light
curve for the nucleus of M87 observed by Chandra between 2000 and
2010 \citep{har09,abr12}. X-ray data of M87 have been taken with the
Advanced CCD Imaging Spectrometer (ACIS) on board the Chandra
satellite \citep{har03,har09,abr12}, which is shown in Figure 3. The
underlying emission is about 0.1 keV/s \citep{har09}, which is
subtracted when we derive the parameters. We identify 18 flares from
the X-ray light curve. The parameters of flares are listed in Table
3, including the start time, end time and energy. We take the error
as 15\% of total energy.

\section{Results}
We present the cumulative energy distribution of X-ray flares in
Figure 4. The number of flares $N(E)dE$ with energy between $E$ and
$E+dE$ can be expressed by
\begin{equation}
N(E)dE\propto E^{-\alpha_E}dE ~~~~ E<E_{\rm max},
\end{equation}
where $\alpha_E$ is the power-law index, and $E_{\rm max}$ is the
cutoff energy. So the cumulative energy distribution is
$N(>E)=a+b[E^{1-\alpha_E}-E_{\rm max}^{1-\alpha_E}]$, where $a$ and
$b$ are two parameters. We use the Markov chain Monte Carlo
technique to obtain the best-fitting parameters. The distribution
shows a flat part at the low energy regime, which may be due to an
incomplete sampling and some selection bias for large energy flares.
Interestingly, Cliver et al. (2012) also interpreted the flatter
slope of solar energetic particle events using selection effects. So
in order to avoid the selection effect, only the cumulative
distribution above the break is fitted. The flattening effect due to
the incomplete sampling is well understood. In Figure 4, we fit the
cumulative distributions in the energy range between the dashed
lines. Because the number of solar flares is very large, we fit the
differential distribution with $\alpha_E=1.65\pm 0.02$ for 11595
solar flares observed by RHESSI \citep{asc11}. The other red curves
in Figure 4 represent the cumulative energy distributions of X-ray
flares with indices $\alpha_E=2.4\pm0.6$, $1.8\pm 0.6$, and $1.6\pm
0.7$ for Swift J1644+57, Sgr A$^*$, and M87, respectively. The
reduced $\chi^2$ of fittings are $\chi^2_r=1.02$, $0.83$, $0.85$,
and $0.75$ for solar flares, Swift J1644+57, Sgr A$^*$, and M87,
respectively. Neilsen et al. (2013) also fitted the energy of Sgr
A$^*$ flares using a cutoff power-law function. The best fitting
slope is $0.9^{+0.8}_{-0.5}$, which is in contrast with our result.
The main reason is that the incomplete sampling at low energy range.
Figure 5 shows the duration-time ($T$) frequency distributions X-ray
flares from black hole systems. The differential distribution can be
expressed as
\begin{equation}
N(T)dT\propto T^{-\alpha_T}dT.
\end{equation}
Because the number of flares from black hole systems is small. We
also use the cumulative distribution, which is the integration of
equation (3). In order to avoid the incomplete sampling problem, the
distribution above the break point is fitted, which is between
dashed lines in Figure 5. The values of $\alpha_T$ are $2.00\pm
0.05$, $1.5\pm 0.6$, $1.9\pm 0.5$, and $2.0\pm0.7$ for solar flares,
Swift J1644+57, Sgr A$^*$, and M87, respectively. The reduced
$\chi^2$ of fittings are $\chi^2_r=0.98$, $0.83$, $0.86$, and $0.74$
for solar flares, Swift J1644+57, Sgr A$^*$, and M87, respectively.
The fitting results are listed in Table 4. Because the flares are
from the same object observed by the same instrument, the
observational bias may be minimal.

Obviously, the typical values of $\alpha_E$ are about $1.6$ and
$2.4$, and those of $\alpha_T$ are about $2.0$ and $1.5$. Some
distributions of flares can be well understood within one physical
framework, i.e., self-organized criticality (SOC). The concept of
SOC was proposed as an explanation for the behavior of the sandpile
model \citep{bak87}. In SOC, subsystems self-organize owing to some
driving force to a critical state at which the ``output" is a series
of ``avalanches" that follow a power-law (fractal) frequency-size
distribution. For solar flares, the statistical power-law
distributions of sizes and durations can be explained by the
universal fractal-diffusive SOC model \citep{asc11,asc12}, while the
underlying physical process of the driver could be attributed to a
magnetic reconnection process. We further discuss this model to
explain the energy and duration-time frequency distributions of
X-ray flares in other systems. For an ensemble of many SOC
avalanches, the relationship between size scale $L$ and duration
time $T$ is $L\propto T^{1/2}$ \citep{asc12}, which has been
originally applied to the Brownian motion. This relationship has
been confirmed by the observations of solar flares \citep{asc12b}.
Meanwhile, under the assumption that the number or occurrence
frequency of avalanches is equally likely throughout the system, the
distribution of size scale $L$ is expressed as $N(L)dL\propto
L^{-S}dL$, where $S=1$, 2 and 3 is the Euclidean dimension. Thus,
the duration frequency distribution of flares is given by
\citep{asc12}
\begin{equation}
N(T)dT\propto T^{-(S+1)/2}dT.
\end{equation}
For $S=3$, the index $\alpha_T=(S+1)/2$ equals $2.0$, which can well
reproduce the duration frequency distributions of X-ray flares of
Sun, Sgr A$^*$, and M87 at the $1\sigma$ confidence level. In addition,
the energy frequency distribution can be expressed as \citep{asc12}
\begin{equation}
N(E)dE\propto E^{-3(S+1)/(S+5)}dE.
\end{equation}
The index $\alpha_E\equiv3(S+1)/(S+5)=1.5$ for $S=3$, which is
consistent with the observed indices of X-ray flares of Sun, Sgr
A$^*$, and M87. The power-law distributions of total energies and
durations are two criteria of a SOC system \citep{asc11}. From our
above statistical analysis, the X-ray flares from Sun, Sgr A$^*$,
and M87 are due to SOC events. The X-ray flares of Sun, Sgr A$^*$,
and M87 correspond to the three-dimensional ($S=3$) case.
Interestingly, Li \& Yuan (2014) also found that the distributions
of Sgr A$^*$ flares, including flux, peak rate and waiting time, can
be explained by three-dimensional fractal-diffusive SOC model. The
best fitting power-law slopes are $\alpha_E=2.4\pm0.6$ and
$\alpha_T=1.5\pm0.6$ for Swift J1644+57. According to the
fractal-diffusive SOC model proposed by Aschwanden (2012a), these
power-law slopes can be marginally explained by the
three-dimensional SOC model. But the difference is up to $0.6$.
Interestingly, the size distributions of stellar flare energies with
power-law slopes in a range of $\alpha_E=2.17\pm0.25$ (Aschwanden
2014). For stellar flares from Kepler mission, the size distribution
for the total sample of 1538 stellar flares shows a power law slope
of $\alpha_E=2.04\pm 0.13$ (Walkowicz et al. 2011; Maehara et al.
2012; Shibayama et al. 2013; Aschwanden 2014). These values are
dramatically consistent with our result, but are steeper than
derived for solar flare energies ($\alpha_E\approx 1.5-1.6$).

The waiting time $\Delta t$ is the time interval between two
successive events in a data set. Figure 6 shows the waiting time
distributions of X-ray flares from the Sun, Swift J1644+57, Sgr
A$^*$ and M87, which suggests that these X-ray flares have similar
waiting-time distributions. The power-law indices of waiting-time
distributions $\alpha_W$ are $2.04\pm 0.03$, $1.8\pm 0.6$, $1.8\pm
0.9$, and $2.9\pm 1.0$ for solar flares, Swift J1644+57, Sgr A$^*$,
and M87, respectively. The corresponding reduced $\chi^2$ are
$\chi^2_r=0.96$, $0.83$, $0.79$, and $0.70$ for solar flares, Swift
J1644+57, Sgr A$^*$, and M87, respectively. The power-law waiting
time distributions of X-ray flares can be explained by
non-stationary Poisson processes \citep{whe98,asc10}. The power-law
index $\alpha_W$ of waiting time distribution is dependent on the
flare rate. For the flare rate distribution
\begin{equation}
f(\lambda) = \lambda^{-1}\exp{\left(-{\lambda /\lambda_0}\right)}
\end{equation}
with flare rate $\lambda=1/\Delta t$ and mean rate $\lambda_0$, the
waiting time distribution can be derived as \citep{asc10}
\begin{equation}\label{wt1}
P(\Delta t) = {\lambda_0 \over (1 + \lambda_0 \Delta t)^2},
\end{equation}
where $\lambda_0$ is the mean rate of flares. For waiting times
$(\Delta t \gg 1/\lambda_0)$, equation \ref{wt1} approaches a
power-law limit $P(\Delta t)\approx (1/\lambda_0)(\Delta t)^{-2}$,
which is consistent with solar flares, GRBs and Sgr A$^*$. Next, we
consider the flare rate distribution with a mean rate $\lambda_0$
\begin{equation}
f(\lambda) =\frac{1}{\lambda_0}\exp{\left(-{\lambda
/\lambda_0}\right)},
\end{equation}
defined in the range of $0<\lambda<\infty$. In this case, the
waiting time distribution can be written as \citep{asc10}
\begin{equation}
P(\Delta t) = \int_0^\infty \left(-{\lambda /\lambda_0}\right)^2
\exp\left[-\frac{\lambda}{\lambda_0}(1+\lambda_0 \Delta t)\right]
d\lambda.
\end{equation}
This equation corresponds to the integral $\int_0^\infty x^2 e^{ax}
dx = -2/a^3$, with $a=-(1+\lambda_0 \Delta t)/\lambda_0$. So the
waiting time distribution is
\begin{equation}
P(\Delta t) = \frac{2\lambda_0}{(1 + \lambda_0 \Delta t)^3}.
\end{equation}
Obviously, the power-law index approaches $-3$ for $\Delta t \gg
1/\lambda_0$, which is remarkably consistent with the X-ray flares
of M87.

\section{Conclusions and discussion} \label{conclusion}
In this paper, we study the statistical properties of X-ray flares
from different objects. The best fitting results for the
distributions of energy, duration and waiting time are shown in
Table 4. The statistical similarities among X-ray flares from the
Sun, GRBs, Swift J1644+57, Sgr A$^*$, and M87, suggest that all of
the X-ray flares have a similar physical origin, i.e., magnetic
reconnection. Our results show that X-ray flares from Sun, Sgr
A$^*$, and M87 can be explained by a three-dimensional SOC model,
implying that relativistic jets may be magnetically dominated. The
power-law slopes of Swift J1644+57 distributions can be marginally
explained in the three-dimensional SOC case. But the difference is
up to $0.6$. X-ray flares in black hole systems may occur in the
following way. The central engine first ejects a relativistic
magnetically-dominated jet \citep{mei01}, and subsequently, blobs
with different velocities in such a jet collide with each
other\citep{yuan12}, triggering magnetic reconnection events.
Finally, relativistic electrons accelerated in the jet, owing to
these events, emit X-ray flares. In addition, our results also show
that SOC events are common phenomena in our universe, from local to
distant objects and from the Sun with $1 M_\odot$ to M87 with a few
$10^9 M_\odot$. This will possibly motivate further studies of
astrophysical SOC events.

\section*{Acknowledgements}
We thank the two referees for detailed and very constructive
suggestions that have allowed us to improve our manuscript. We thank
M. J. Aschwanden, L. A. Balona, P. F. Chen, M. D. Ding, Y. F. Huang,
X. Y. Wang, F. Yuan, and S. N. Zhang for their discussions. This
work is supported by the National Basic Research Program of China
(973 Program, grant No. 2014CB845800), the National Natural Science
Foundation of China (grants 11422325, 11373022, 11103007, and
11033002), the Excellent Youth Foundation of Jiangsu Province
(BK20140016), and the Program for New Century Excellent Talents in
University (grant No. NCET-13-0279). This work made use of data
supplied by the UK Swift Science Data Center at the University of
Leicester.

\clearpage
\begin{longtable}{ccccccccccc}
\multicolumn{11}{l}{{\bf Table~1.} The measured parameters of X-ray flares of Swift J1644+57.} \\ \\
%\multicolumn{11}{l}{{\bf Table~S1.} The measured parameters of X-ray flares of GRBs.} \\
\hline \hline $t_{\rm start}^{\rm a}$ & $t_{\rm end}$$^{\rm b}$ & $S_F^{\rm c}$ & $\sigma_{S_F}^d$ & $E_{\rm iso}^{\rm e}$ \\
 (s) & (s) & ($10^{-8}$\,erg\,cm$^{-2}$) & ($10^{-8}$\,erg\,cm$^{-2}$) & ($10^{50}$\,ergs) \\
 \hline
2019.2 & 2814.3 & 36.4 & 8.2 & 1.10 \\
17750.2 & 27267.8 & 17.7 & 5.3 & 0.53 \\
33099.6 & 50059.5 & 129.5 & 25.2 & 3.90 \\
45489.5 & 49643.7 & 73.7 & 15.0 & 2.22 \\
39025.2 & 1.71E5 & 128.7 & 32.3 & 3.88 \\
75411.5 & 90710.2 & 23.6 & 6.1 & 0.71 \\
1.10E5 & 1.12E5 & 13.6 & 3.2 & 0.41 \\
1.108E5 & 1.116E5 & 3.37 & 1.4 & 0.102 \\
1.109E5 & 1.469E5 & 88.2 & 17.6 & 2.66 \\
1.121E5 & 1.129E5 & 6.48 & 1.5 & 0.195 \\
1.331E5 & 1.602E5 & 132.6 & 27.3 & 3.99 \\
1.393E5 & 1.422E5 & 12.39 & 2.4 & 0.373 \\
1.397E5 & 1.401E5 & 6.89 & 1.4 & 0.208 \\
1.391E5 & 1.445E5 & 142.5 & 35.0 & 4.30 \\
1.451E5 & 1.472E5 & 9.91 & 2.4 & 0.299 \\
1.455E5 & 1.476E5 & 16.18 & 4.1 & 0.488 \\
1.729E5 & 2.264E5 & 57.81 & 12.3 & 1.742 \\
1.738E5 & 1.894E5 & 15.89 & 3.5 & 0.479 \\
1.852E5 & 1.912E5 & 6.00 & 1.4 & 0.181 \\
1.869E5 & 1.947E5 & 147.5 & 37.9 &  4.45 \\
2.028E5 & 2.216E5 & 24.50 & 5.3 & 0.738 \\
2.085E5 & 2.195E5 & 95.87 & 21.3 & 2.89  \\
2.154E5 & 3.516E5 & 173.1 & 38.2 & 5.22 \\
2.666E5 & 3.017E5 & 83.14 & 13.4 & 2.51 \\
2.955E5 & 4.198E6 & 3496.2 & 450.1 & 105.4 \\
3.172E5 & 3.667E5 & 162.4 & 31.7 & 4.89 \\
3.901E5 & 5.377E5 & 326.0 & 50.6 & 9.83 \\
5.236E5 & 9.966E5 & 495.6 & 89.8 & 14.94 \\
6.020E5 & 6.109E5 & 15.24 & 3.2 & 0.459 \\
6.184E5 & 6.291E5 & 24.61 & 4.3 & 0.742 \\
7.642E5 & 9.235E5 & 545.5 & 97.3 & 16.44 \\
9.557E5 & 1.562E6 & 1444.3 & 201.5 & 43.53 \\
1.204E6 & 1.333E6 & 620.8 & 80.4 & 18.71 \\
1.309E6 & 1.394E6 & 116.7 & 20.0 & 3.52 \\
1.363E6 & 1.447E6 & 85.20 & 25.8 & 2.57 \\
1.459E6 & 1.461E6 & 3.79 & 0.9 & 0.114 \\
1.463E6 & 1.543E6 & 115.9 & 20.3 & 3.49 \\
1.436E6 & 1.597E6 & 242.1 & 35.2 & 7.30 \\
1.540E6 & 1.965E6 & 996.8 & 130.6 & 30.04 \\
1.856E6 & 2.059E6 & 247.8 & 37.8 & 7.47 \\
2.067E6 & 2.307E6 & 294.6 & 40.0 & 8.88 \\
2.280E6 & 3.581E6 & 428.0 & 89.7 & 12.90 \\
2.342E6 & 2.524E6 & 60.30 & 15.3 & 1.82 \\
2.442E6 & 2.562E6 & 255.9 & 63.4 & 7.71 \\
2.597E6 & 3.481E6 & 378.7 & 78.2 & 11.41 \\
2.780E6 & 2.875E6 & 108.4 & 30.8 & 3.27 \\
2.942E6 & 3.727E6 & 377.3 & 64.2 & 11.37 \\
3.188E6 & 3.246E6 & 100.9 & 20.4 & 3.04 \\
3.188E6 & 3.691E6 & 1289.9 & 300.2 & 38.87 \\
3.599E6 & 3.986E6 & 228.8 & 46.2 & 6.89 \\
4.007E6 & 4.360E6 & 109.2 & 18.9 & 3.29 \\
4.335E6 & 4.536E6 & 159.7 & 31.8 & 4.81 \\
4.404E6 & 4.987E6 & 353.2 & 83.2 & 10.65 \\
5.336E6 & 6.246E6 & 291.4 & 50.7 & 8.78 \\
6.065E6 & 6.195E6 & 19.35 & 5.8 & 0.583 \\
6.064E6 & 7.228E6 & 437.7 & 67.4 & 13.19 \\
7.142E6 & 7.314E6 & 20.60 & 5.2 & 0.62 \\
7.029E6 & 7.979E6 & 257.0 & 63.8 & 7.75 \\
7.942E6 & 8.265E6 & 64.7 & 13.5 & 1.95 \\
8.359E6 & 8.562E6 & 23.32 & 6.1 & 0.70 \\
8.293E6 & 1.250E7 & 344.3 & 69.5 & 10.38 \\
9.326E6 & 9.869E6 & 39.2 & 7.3 & 1.18 \\
9.329E6 & 1.437E7 & 366.7 & 73.8 & 11.05 \\
9.773E6 & 1.295E7 & 870.3 & 143.8 & 26.23 \\
1.297E7 & 2.511E7 & 199.1 & 45.2 & 6.00 \\
1.426E7 & 1.712E7 & 516.2 & 34.9 & 15.56 \\
1.635E7 & 3.996E7 & 607.5 & 130.2 & 18.31 \\
2.528E7 & 9.447E7 & 343.2 & 65.2 & 10.34\\
\hline
\end{longtable}
Note: (a) The start time of a flare. (b) The end time of a flare.
(c) The fluence of flare. (d) The error of fluence.  (e) The flare
energy.

\clearpage
\begin{table*}
\centering \caption{The parameters of X-ray flares of Sgr A$^*$.
Note: (a) The start time of a flare in units of Modified Julian Date
(MJD). (b) The end time of flare in units of MJD. (c) The fluence of
flares. (d) The duration of flares.}
\begin{tabular}{ccccc}
\hline\hline $t_{\rm start}^{\rm a}$ & $t_{\rm end}$$^{\rm b}$ & $S_F^{\rm c}$ & T$^{\rm d}$ & $L_{2-10}^{\rm unabs}$ \\
 (MJD) & (MJD) & (cts) & (s) & ($10^{34}$\,erg~s$^{-1}$) \\
 \hline
55966.433 & 55966.464 & $33^{+12}_{-11}$ &2600 & 1.7 \\
55966.603 & 55966.666 & $706^{+46}_{-44}$ & 5450 &19.2 \\
56006.486 & 56006.493 & $32^{+11}_{-9}$ & 600 & 7.4 \\
56006.524 & 56006.540 & $40^{+12}_{-10}$ &1350 & 4.1 \\
56006.580 & 56006.599 & $49^{+13}_{-12}$ &1600 & 4.2 \\
56006.678 & 56006.690 & $49^{+13}_{-14}$ &950 & 7.1 \\
56048.510 & 56048.548 & $59^{+16}_{-14}$ &3250 &2.5 \\
56048.679 & 56048.693 & $15^{+8}_{-7}$ &1200 & 1.7 \\
56054.107 & 56054.154 & $49^{+15}_{-13}$ &4050 & 1.6 \\
56058.687 & 56058.705 & $24^{+10}_{-11}$ &1600 & 2.0 \\
56059.006 & 56059.044 & $33^{+15}_{-11}$ &3250 & 1.4 \\
56059.314 & 56059.329 & $21^{+10}_{-8}$ & 1250 & 2.3 \\
56060.127 & 56060.168 & $124^{+21}_{-19}$ &3500 & 4.9 \\
56067.863 & 56067.888 & $102^{+18}_{-17}$ &2150 & 6.6 \\
56122.650 & 56122.656 & $ 8^{+6}_{-4}$ & 500 &2.3 \\
56126.979 & 56127.038 & $58^{+16}_{-15}$ & 5100 & 1.6 \\
56127.172 & 56127.202 & $26^{+11}_{-9}$ & 2550 & 1.4 \\
56128.549 & 56128.553 & $29^{+10}_{-9}$ & 400& 10.2 \\
56130.182 & 56130.225 & $101^{+19}_{-17}$ &3700 & 3.8 \\
56130.906 & 56130.921 & $46\pm13$ & 1300  & 4.8 \\
56131.494 & 56131.585 & $119^{+23}_{-21}$ & 7800 & 2.1 \\
56132.385 & 56132.399 & $38\pm12$  & 1150 & 4.5 \\
56133.512 & 56133.521 & $14^{+8}_{-7}$ &750 & 2.6 \\
56133.997 & 56134.042 & $251^{+28}_{-26}$ & 3950 & 8.9 \\
56139.368 & 56139.417 & $166^{+24}_{-22}$ & 4250 & 5.4 \\
56141.009 & 56141.035 & $135^{+21}_{-23}$ & 2250 & 8.4 \\
56143.314 & 56143.332 & $58\pm15$ & 1550 & 5.1 \\
56144.321 & 56144.363 & $33^{+14}_{-12}$ & 3600 & 1.2 \\
56147.131 & 56147.151 & $27^{+11}_{-10}$ &1750 & 2.1 \\
56207.174 & 56207.194 & $30^{+11}_{-10}$ & 1700 & 2.4 \\
56208.187 & 56208.222 & $54^{+15}_{-13}$ & 2950 & 2.5 \\
56216.239 & 56216.248 & $58^{+14}_{-12}$ & 750 & 10.7 \\
56217.094 & 56217.098 & $15^{+8}_{-6}$ & 400 & 5.4 \\
56217.816 & 56217.884 & $372^{+34}_{-32}$ & 5900 & 8.9 \\
56223.384 & 56223.464 & $193^{+26}_{-24}$ &6900 & 3.8 \\
56225.230 & 56225.263 & $63^{+16}_{-14}$ & 2800 & 3.1 \\
 56230.288 & 56230.362 & $74^{+19}_{-18}$ & 6350 & 1.6 \\
56230.724 & 56230.734 & $13^{+8}_{-7}$ & 900 & 2.0 \\
56231.566 & 56231.592 & $54^{+14}_{-13}$ & 2250  & 3.3\\
\hline
\end{tabular}
\end{table*}

\clearpage
\begin{table*}
\centering \caption{The parameters of X-ray flares of M87. Note: (a)
The start time of a flare. (b) The end time of a flare. (c) The
flare energy.}
\begin{tabular}{cccc}
\hline\hline $t_{\rm start}^{\rm a}$ & $t_{\rm end}$$^{\rm b}$ & $E^{\rm c}$ \\
(year) & (year) & ($10^{47}$ ergs) \\
\hline
1999.3169 &  2003.4192 &  9.06\\
2001.3497 &  2004.4083 &  6.05\\
2002.3079 &  2007.8254 &  2.02\\
2002.7067 &  2003.8173 &  2.05\\
2003.3964 &  2004.1254 &  1.92\\
2003.6656 &  2004.8537 &  3.53\\
2004.3156 &  2005.2375 &  1.00\\
2004.4449 &  2006.3123 &  3.72\\
2005.1603 &  2005.3801 &  3.58\\
2005.2945 &  2005.6828 &  1.99\\
2005.4406 &  2006.4997 &  0.11\\
2006.2941 &  2007.4452 &  0.40\\
2006.9754 &  2007.9632 &  6.83\\
2006.9998 &  2007.6756 &  1.41\\
2007.4370 &  2008.4702 &  2.17\\
2008.0289 &  2010.7990 &  0.11\\
2008.1739 &  2009.0230 &  10.1\\
2008.1969 &  2010.3154 &  2.35\\
\hline
\end{tabular}
\end{table*}

\clearpage
\begin{table}
%\begin{center}
\caption{The properties of X-ray flares.}
\begin{tabular}{ccccccc}
\hline \hline
Source & flare energy & number of flares & $\alpha_E$ & $\alpha_T$ & $\alpha_W$ & dimension \\
& (ergs)&  & & & & \\
\hline
Sun &  $10^{26}$-$10^{32}$ & 11595 & $1.65\pm0.02$ & $2.00\pm0.05$ & $2.04\pm0.03$ & $S=3$ \\
GRBs &  $10^{48}$-$10^{52}$ & 83 & $1.1\pm0.2$ & $1.1\pm0.2$ & $1.8\pm0.2$ & $S=1$ \\
Swift J1644+57 &  $10^{48}$-$10^{52}$ & 68 & $2.4\pm0.6$ & $1.5\pm0.6$ & $1.8\pm0.6$ & $S=3?$ \\
Sgr A$^*$ &  $10^{37}$-$10^{40}$ & 39 & $1.8\pm0.6$ & $1.9\pm0.5$ & $1.8\pm0.9$ & $S=3$ \\
M87 &  $10^{46}$-$10^{48}$ & 18 & $1.6\pm0.7$ & $2.0\pm0.7$ & $2.9\pm1.0$ & $S=3$ \\
\hline
\end{tabular}
Note: $\alpha_E$, $\alpha_T$, and $\alpha_W$ are the power-law
indices of the frequency distributions of flare energies, durations,
and waiting times, respectively.
%\end{center}
\end{table}

\clearpage

\begin{figure}
\includegraphics[width=\textwidth]{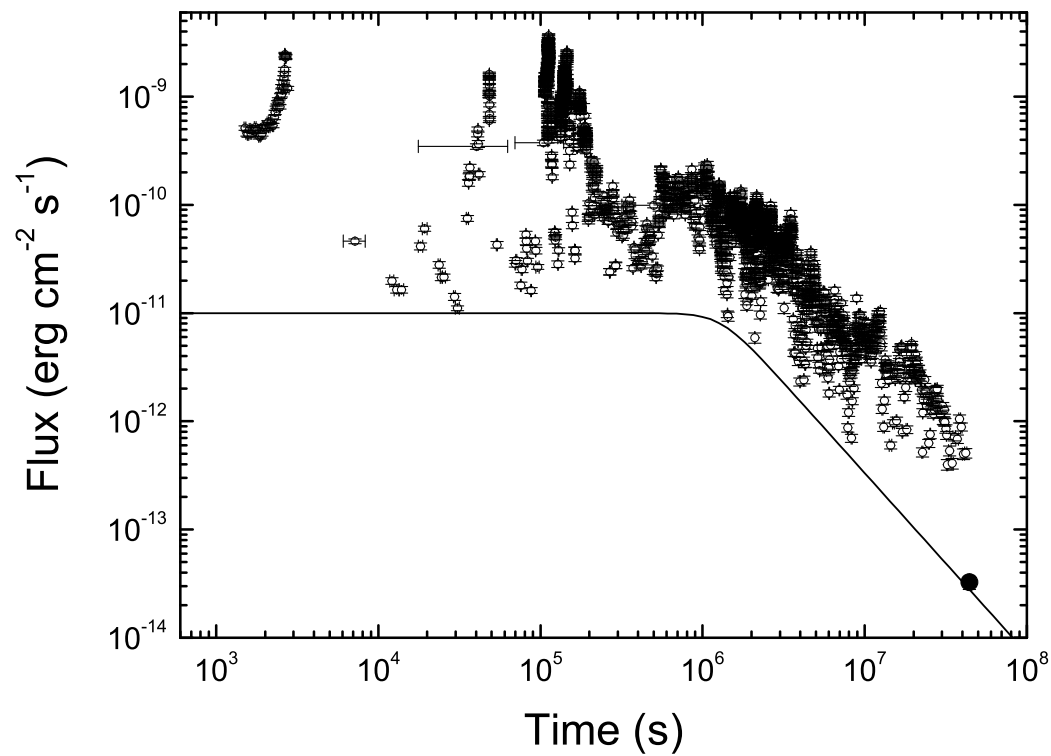}
\caption{\label{Fig1} X-ray light curve of Swift J1644+57 from
Swift/XRT (open circles) and a late-time Chandra observation (filled
circle). The black line is the underlying continuum emission with a
constant flux at $t<15$ days and $F_X\propto t^{-5/3}$ at $t>15$
days. After the relativistic jet shuts off, the underlying continuum
emission is consistent with the Chandra observation. The underlying
continuum emission has been subtracted when we fit the parameters of
flares.}
\end{figure}

\begin{figure}
\includegraphics[width=\textwidth]{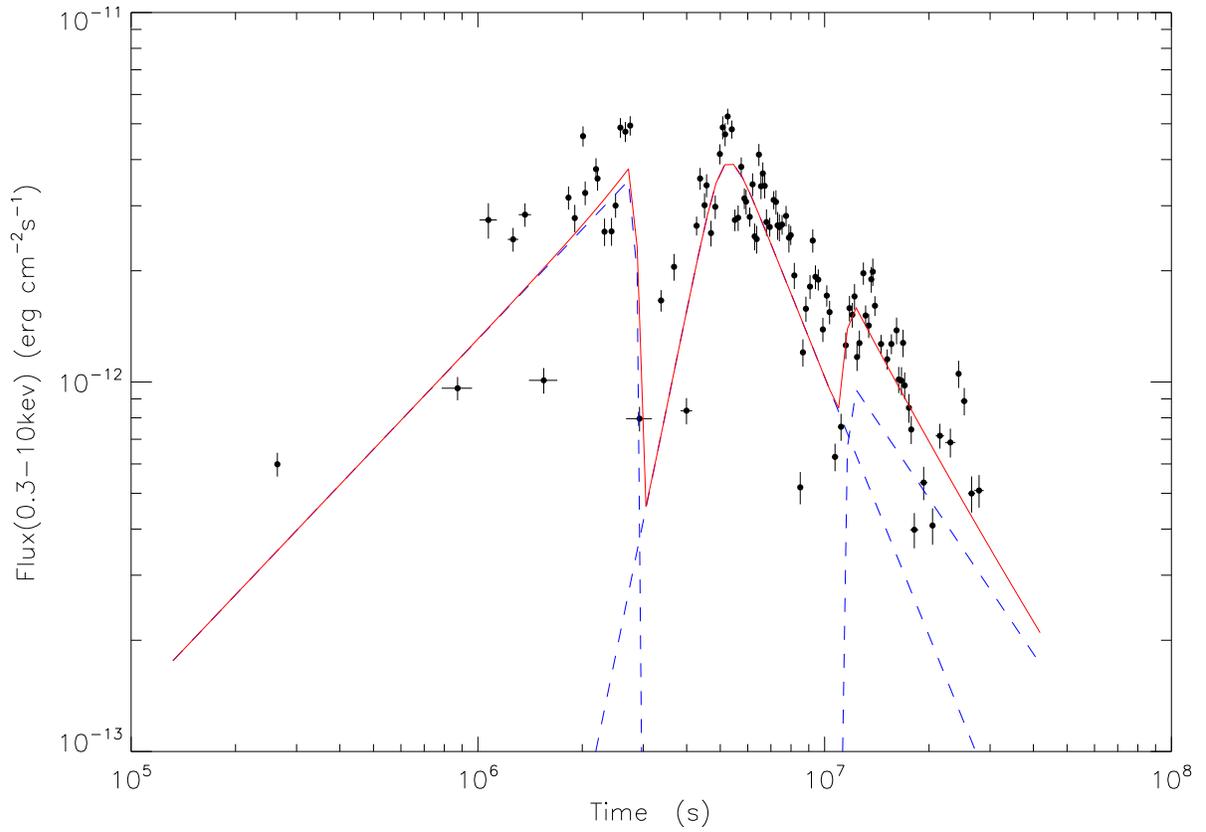}
\caption{\label{Fig2} The best fit of three flares of Swift J1644+57
with three power-law functions after the subtraction of the
underlying continuum. The start time is the start time minus
$1.42\times 10^7$ s. The red line shows the total best fit, and the
dash lines show the best fits for individual flares.}
\end{figure}

\begin{figure}
\includegraphics[width=\textwidth]{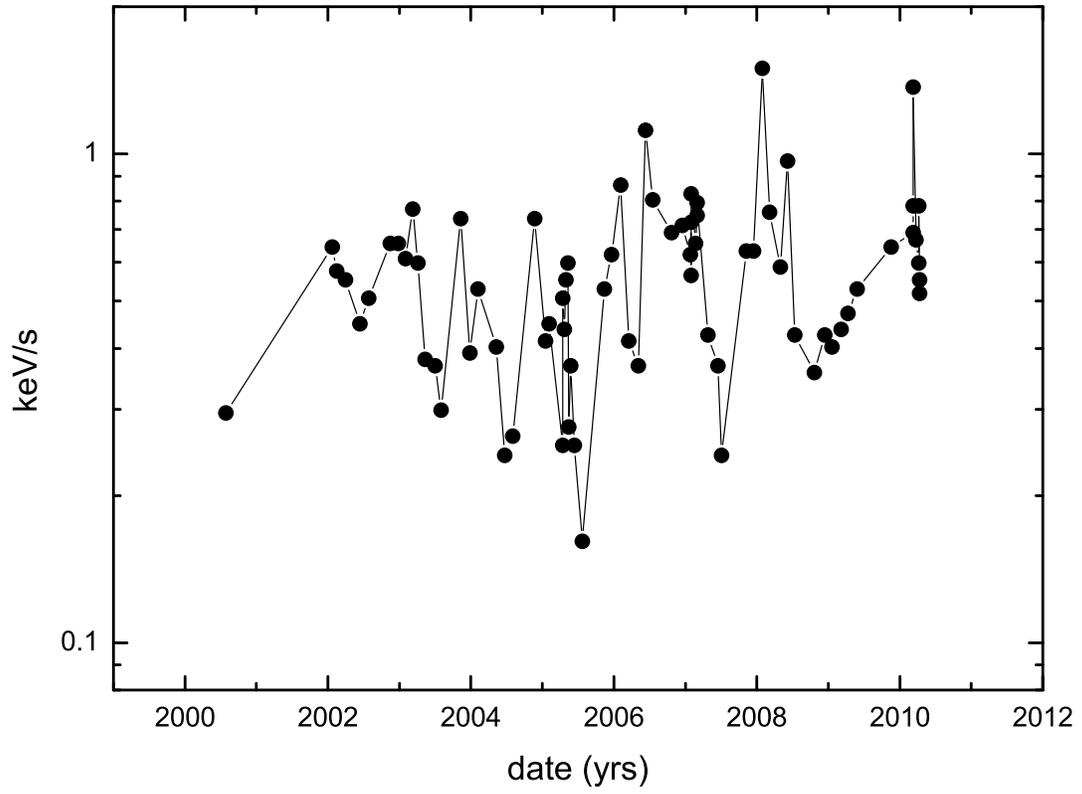}
\caption{\label{Fig3} The X-ray light curve for the nucleus of M87
observed by Chandra.}
\end{figure}

\begin{figure}
\centering
\includegraphics[width=\textwidth]{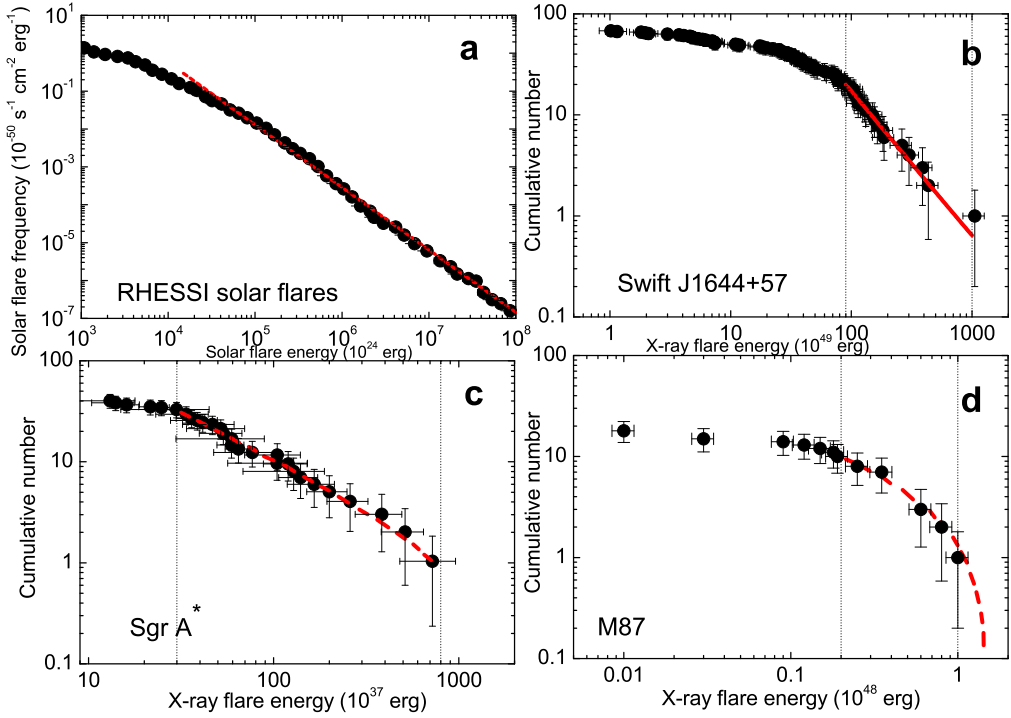}
\caption{\label{Fig4} Cumulative size distributions of energy for
X-ray flares. (\textbf{a}) The black dots are 11595 X-ray flares of
the Sun observed by RHESSI. The red curve gives the differential
energy distribution $dN/dE\propto E^{-\alpha_E}$, with $\alpha_E =
1.65\pm 0.02$. (\textbf{b}) 68 X-ray flares from Swift J1644+57 are
shown as black dots. The red curve gives the cumulative energy
distribution $N(>E)=a+b[E^{1-\alpha_E}-E_{\rm max}^{1-\alpha_E}]$
with $\alpha_E=2.4\pm 0.6$. The fitting range is between the two
vertical dash lines. (\textbf{c}) 39 X-ray flares from Sgr A$^*$
observed by Chandra are shown as black dots. The red curve gives the
best fit with $\alpha_E=1.8\pm 0.6$. The fitting range is between
the two vertical dash lines. (\textbf{d}) 18 X-ray flares of M87
observed by Chandra are shown as black dots. The red curve gives
$\alpha_E=1.6\pm 0.7$. The fitting range is between the two vertical
dash lines. }
\end{figure}

\begin{figure}
\centering
\includegraphics[width=\textwidth]{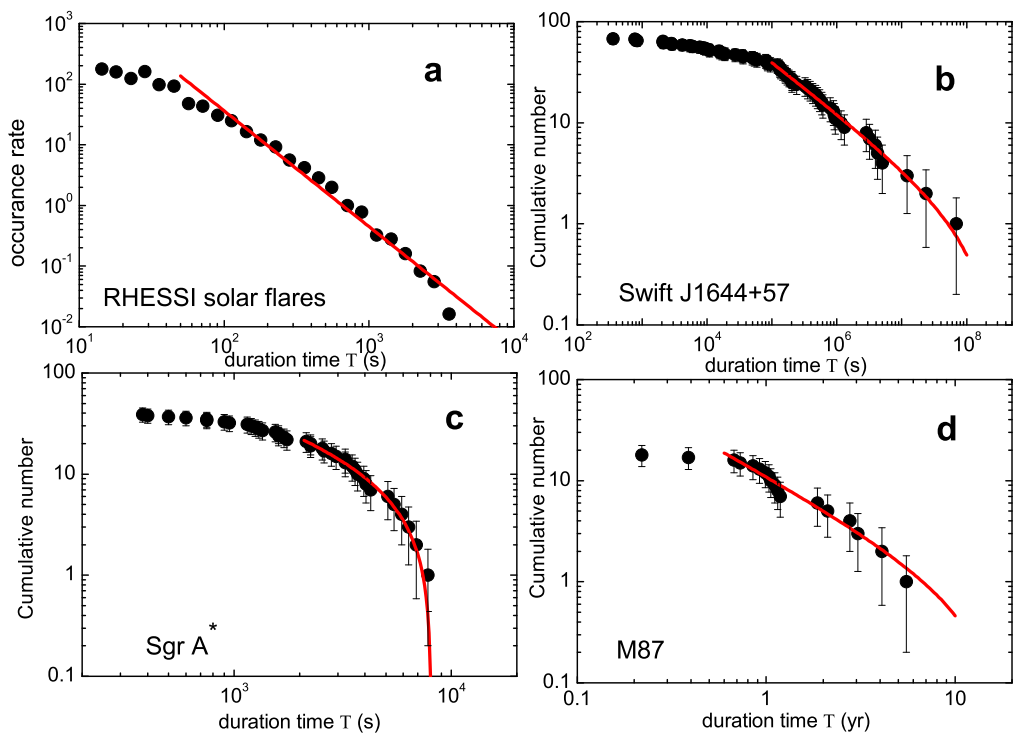}
\caption{\label{Fig5} Cumulative size distributions of duration
time. (\textbf{a}-\textbf{d}) The best-fit values of $\alpha_T$ are
$2.00\pm 0.05$, $1.5\pm 0.6$, $1.9\pm 0.5$, and $2.0\pm0.7$ for
solar flares, Swift J1644+57, Sgr A$^*$, and M87, respectively. The
fitting range is between the two vertical dash lines.}
\end{figure}

\begin{figure}
\centering
\includegraphics[width=\textwidth]{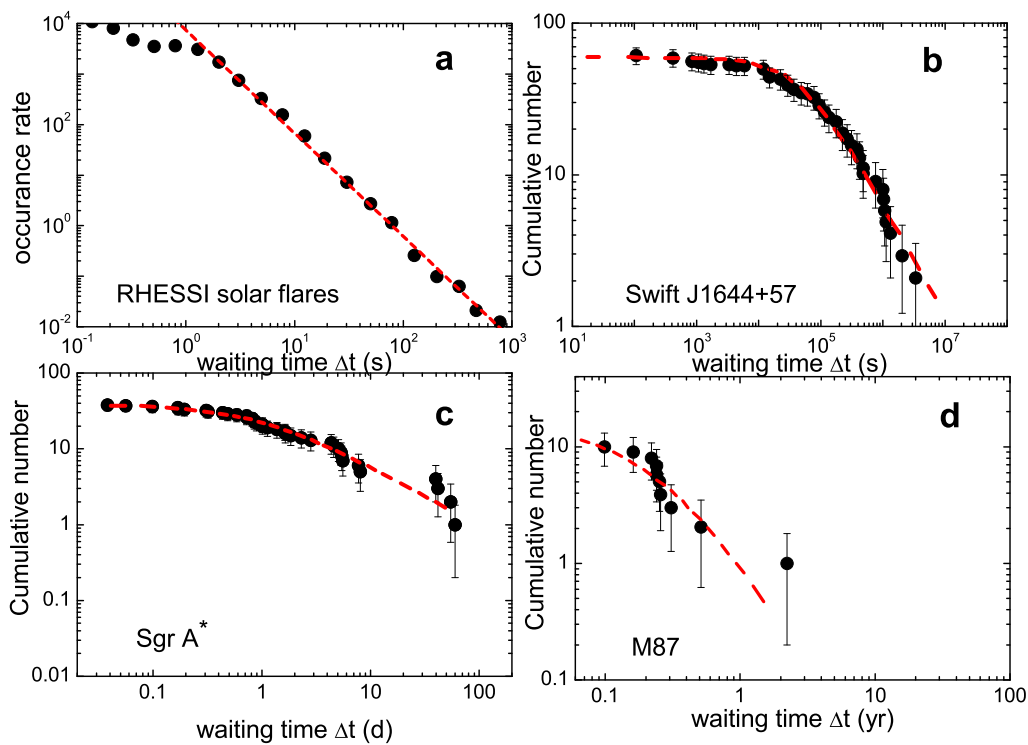}
\caption{\label{Fig6} Cumulative size distributions of waiting time.
(\textbf{a}-\textbf{d}) The best-fit power-law indices $\alpha_W$
are $2.04\pm 0.03$, $1.8\pm 0.6$, $1.8\pm 0.9$, and $2.9\pm1.0$ for
solar flares, Swift J1644+57, Sgr A$^*$, and M87, respectively.}
\end{figure}

\end{document}